\documentclass[letterpaper]{article} 
\usepackage[]{aaai25}  
\usepackage{times}  
\usepackage{helvet}  
\usepackage{courier}  
\usepackage[hyphens]{url}  
\usepackage{graphicx} 
\usepackage{natbib}  
\usepackage{caption} 
\usepackage{algorithm}
\usepackage{algorithmic}

\usepackage[table]{xcolor}

\definecolor{stage1}{HTML}{FEF4DA}
\definecolor{stage2}{HTML}{C5F9F4}
\definecolor{data1}{HTML}{FFE0CA}
\definecolor{data2}{HTML}{F3EEFC}
\definecolor{actor1}{HTML}{FFFFD2}
\definecolor{actor2}{HTML}{FFE1F8}

\usepackage{newfloat}
\usepackage{listings}
\DeclareCaptionStyle{ruled}{labelfont=normalfont,labelsep=colon,strut=off} 
\floatstyle{ruled}
\newfloat{listing}{tb}{lst}{}
\floatname{listing}{Listing}
\usepackage{array}
\usepackage{longtable}
\usepackage{booktabs} 

\title{Disclosure and Evaluation as Fairness Interventions for General-Purpose AI}
\author{
    Vyoma Raman\textsuperscript{\rm 1,2},
    Judy Hanwen Shen\textsuperscript{\rm 1},
    Andy K. Zhang\textsuperscript{\rm 1},
    Lindsey Gailmard\textsuperscript{\rm 1},
    Rishi Bommasani\textsuperscript{\rm 1},
    Daniel E. Ho\equalcontrib\textsuperscript{\rm 1},
    Angelina Wang\equalcontrib\textsuperscript{\rm 2}
}
\affiliations{
    \textsuperscript{\rm 1}Stanford University \\
    \textsuperscript{\rm 2}Cornell Tech \\
    \vspace{0.5em}
    vr359@cornell.edu
    
}

\usepackage{bibentry}

\begin{document}
%
\maketitle
\begin{abstract}
Despite conflicting definitions and conceptions of fairness, AI fairness researchers broadly agree that fairness is context-specific. However, when faced with general-purpose AI, which by definition serves a range of contexts, how should we think about fairness? We argue that while we cannot be prescriptive about what constitutes fair outcomes, we can specify the processes that different stakeholders should follow in service of fairness. Specifically, we consider the obligations of two major groups: system providers and system deployers. While system providers are natural candidates for regulatory attention, the current state of AI understanding offers limited insight into how upstream factors translate into downstream fairness impacts. Thus, we recommend that providers invest in evaluative research studying how model development decisions influence fairness and disclose whom they are serving their models to, or at the very least, reveal sufficient information for external researchers to conduct such research. On the other hand, system deployers are closer to real-world contexts and can leverage their proximity to end users to address fairness harms in different ways. Here, we argue they should responsibly disclose information about users and personalization and conduct rigorous evaluations across different levels of fairness. Overall, instead of focusing on enforcing fairness outcomes, we prioritize intentional information-gathering by system providers and deployers that can facilitate later context-aware action. This allows us to be specific and concrete about the processes even while the contexts remain unknown. Ultimately, this approach can sharpen how we distribute fairness responsibilities and inform more fluid, context-sensitive interventions as AI continues to advance.
\end{abstract}

\section{Introduction}

As AI is applied to new domains and deployed in new contexts, the potential for fairness-related harms grows significantly. These harms do not remain isolated; rather, biases can cascade and amplify inequities across different levels of harm, underscoring the complexity and urgency of addressing fairness in AI systems. A key refrain in fairness literature is that researchers cannot ``treat fairness and justice... separate from a social context'' \cite{selbst2019sociotechnical}. This principle informs everything from which metrics to choose to which social groups to consider to how to determine and represent relevant factors of individuals. However, the emergence of general-purpose AI models (GPAI), often referred to as foundation models~\cite{bommasani2021foundation}, complicates this approach. GPAI is characterized by its applicability to a wide range of tasks, many of which may be unforeseen at the time of development. Since context is lacking, we avoid prescribing specific fairness outcomes (e.g., that a decision-making model has selection rates that are at least 80\% of each other). Instead, we advocate for the gathering of information---specifically, disclosure of contextual and supply chain information and evaluation of AI systems under different design decisions and inputs---that can be used to understand and improve fairness across different levels. This information can facilitate later context-aware action.

When considering a general-purpose AI system, dividing the analysis of fairness harms across different levels offers a systematic approach to trace how inequities emerge and escalate. Research illustrates how fairness harms occur across three levels of AI deployment: the model level, system level, and society level \cite{sureshframework}. At the model level, biases in training data or algorithmic design can produce disparities across demographic groups. For example, melanoma detection models often demonstrate higher accuracy for lighter skin tones \cite{daneshjou2022disparities, montoya2025towards}. At the system level, these biases can intensify when the model is integrated into decision support systems. When AI is deployed in domains like policing or hiring, outputs are actively recontextualized by human decision-makers who hold different levels of skepticism and agency that affect potential inequities \cite{brayne2021technologies, kiviat2018art}. At the society level, the compounded effects of biased decision-making tools can exacerbate structural inequities. For instance, in medical diagnosis, persistent inequalities in critical predictive attributes like race or health cost \cite{obermeyer2019dissecting, eneanya2019reconsidering} can combine with underperforming AI systems. This may result in reduced access to timely and accurate treatment for underserved populations, further entrenching inequities in healthcare delivery. At the system and society level, the context where the AI is being applied shapes its ultimate effects. Together, these levels highlight not only how harms escalate but also where different kinds of disclosure and evaluation are needed to understand and mitigate inequities.

\begin{figure*}
    \centering
    \includegraphics[width=0.97\linewidth]{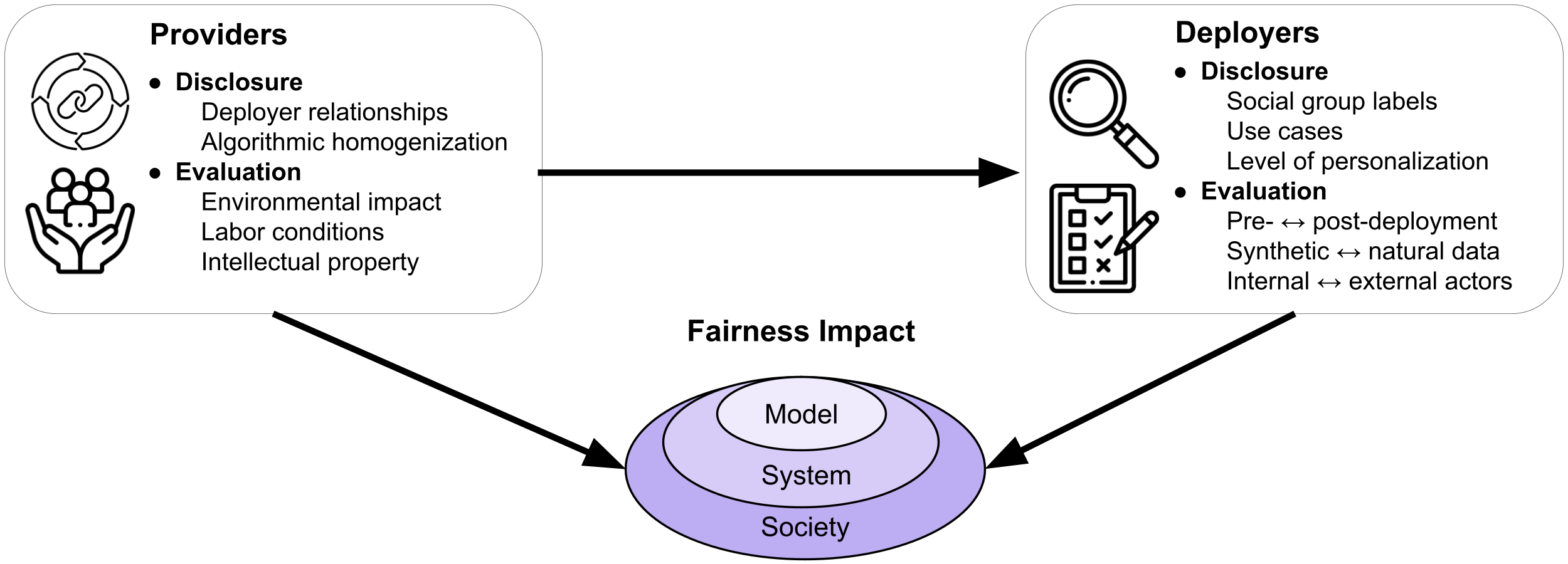}
    \caption{We recommend fairness-related obligations involving information-gathering for providers and deployers to help us analyze and mitigate harms at the model, system, and society level.}
    \label{fig:summary}
\end{figure*}

Within the AI lifecycle, we distinguish between two primary roles: system \textit{providers} and system \textit{deployers}.
System providers are entities that make GPAI models accessible to others, either by distributing them directly or by exposing them through their own interfaces. This definition includes those who release pretrained models (e.g., Meta with Llama), provide platform access to models via APIs (e.g., OpenAI, Anthropic), or facilitate model sharing through hubs (e.g., Hugging Face, Together AI). 
What qualifies an entity to be a provider is not whether they developed or fine-tuned the underlying model, but rather that they made it available for external use in a general-purpose capacity, either in its original or modified form. 
The obligations we place upon system providers can also be satisfied by those who develop the models.\footnote{It may be worth taking different approaches to defining this group as new research grows our understanding of how developer decisions (e.g., training data curation) affect downstream tasks.} System deployers, by contrast, are entities that integrate a model into end-user-facing or internal applications. This includes incorporating GPAI into customer-facing tools like chatbots (e.g., Klarna's AI assistant using OpenAI's models), embedding models in productivity platforms (e.g., Microsoft's Copilot), or leveraging them internally for tasks such as document summarization, code generation, or predictive analytics. These definitions largely align with the EU AI Act~\cite{eu2024aiact}.
We examine fairness-related obligations for system providers with limited foresight into post-deployment modifications and uses, and for deployers who may not fully anticipate eventual applications. We delineate the respective responsibilities of the providers and deployers of AI systems and how they address different levels of impact, as shown in Figure \ref{fig:summary}.

Despite growing attention to fairness in generative AI, existing frameworks often struggle to assess and mitigate harms across diverse and shifting contexts, particularly when systems are designed for general-purpose use and deployed in unforeseen ways. Much of the current scholarship on generative AI focuses on representational harms, examining how groups or individuals are misrepresented in AI outputs \cite{barocas2017problem, Katzman_Wang_Scheuerman_Blodgett_Laird_Wallach_Barocas_2023, ircai2024challenging, teo2023Measuring, Ghosh_Lutz_Caliskan_2024, hofmann2023ai}. While these studies include sophisticated analyses of subtle and structural biases, they often emphasize model-level attributes (e.g., linguistic or visual patterns), which may not fully capture how these representations function in real-world deployments or interact with broader sociotechnical systems. This gap becomes more pronounced in GPAI, where the lack of predefined use cases makes it harder to anticipate how harms might emerge across different applications \cite{ferrara2024fairness, gallegos-etal-2024-bias, weidinger2022taxonomy, cohen2025two}. Depending on the context, harms may extend beyond biased representations to include issues such as information leakage, privacy violations, and the spread of disinformation or toxic content \cite{weidinger2022taxonomy, Xiang2024fairness, Kearns2024responsible}. These cascading harms demand fairness frameworks sensitive to both representational concerns and deployment contexts.

Given the seeming paradox of fairness requiring contextual understanding, and general-purpose AI lacking context, we emphasize forms of systematic information-gathering to support more subsequent context-sensitive responses. Specifically, disclosure is needed about users and supply chain relationships, and evaluation is needed for the different levels of fairness risks an AI system could pose and how the behavior of a system is affected by factors including fine-tuning datasets. Importantly, we advocate for evaluation that surfaces biased representations or disparities in performance metrics \textit{in addition to} research that investigates how to apply this information to effectively mitigate fairness issues at different levels. Information-gathering can itself act as a fairness intervention by increasing transparency, enabling accountability, and creating market pressure for companies to adopt more ethical practices \cite{wang2024strategies, raji2019actionable}. More importantly, it lays the groundwork for adaptive governance by allowing researchers, policymakers, and civil society actors to better respond when fairness harms emerge, accumulate, or shift over time. By collecting and analyzing data from across the AI development and deployment pipeline, we can develop more targeted, flexible interventions that are responsive to the dynamic nature of GPAI systems and carefully scoped to avoid undue burdens on smaller actors.

\subsection{Our Work}
In this work, we propose an approach to addressing fairness-related harms in GPAI, emphasizing the role of gathering information on GPAI and its deployment contexts as a critical intervention strategy in contexts where future use cases and model capabilities are uncertain. Our paper contains the following sections:

\begin{itemize}
\item \textbf{Expanding the Scope of Fairness Interventions:} We draw on prior work that has broadened the scope of fairness beyond the model level to focus on the system and society level. We argue for increased information-gathering to better understand the scope of risk at these higher levels, as well as how to intervene.
\item \textbf{Provider Obligations:} We articulate fairness-related responsibilities for system providers, including supporting research ecosystems and increasing transparency around model recipients and provenance. In doing so, we challenge the common intuition that fairness can be ensured through dataset curation or universal disparity mitigation.
\item \textbf{Deployer Obligations:} We outline deployer obligations under broad deployment conditions, emphasizing increased transparency and diverse evaluations that can better characterize use.
\item \textbf{Operationalizing Fairness for Regulatory Contexts:} We offer recommendations on setting guidelines for which models and systems are subject to particular types of regulation, so as not to overly burden less-resourced companies, and we discuss alignment with ongoing regulatory developments.
\end{itemize}

\section{Expanding the Scope of Fairness Interventions}

We contextualize information-gathering as a critical mechanism for surfacing fairness harms across the model, system, and society levels of GPAI systems. First, we describe how information-gathering can operate across three levels of fairness. Next, we examine how current legislation addresses and distributes responsibility for fairness in GPAI across actors and argue for a reallocation of obligations. Finally, we describe the role of information-gathering in enhancing fairness interventions.

\subsection{Levels of Fairness}

As AI systems diffuse across sectors and are adapted to diverse and often unforeseen applications, evaluating fairness solely at the level of model outputs is insufficient. In contexts where the eventual uses of general-purpose AI are uncertain, disaggregating harms into the model, system, and society levels provides a structured way to capture how inequities emerge, escalate, and compound. While prior work has emphasized the importance of moving beyond predictive disparities \cite{green2018myth, selbst2019sociotechnical, wang2022representational}, these broader harms remain under-measured, particularly when GPAI is repurposed in settings where existing benchmarks fail to reflect real risks. A multi-level framework allows fairness analyses to extend from technical design choices to institutional workflows and socio-economic structures, ensuring that information-gathering practices can anticipate harms wherever they arise.

\paragraph{Model Level.} Model-level fairness analyses quantify disparities in outputs and predictions that result from datasets, model architectures, and training practices. These studies focus particularly on representational and allocational harms \cite{barocas2017problem}, such as racial disparities in medical diagnostic models or gender biases in hiring algorithms. For instance, models for melanoma detection are more accurate for lighter skin tones due to the overrepresentation of lighter-skinned patients in training datasets \cite{daneshjou2022disparities, montoya2025towards}. Similarly, automated speech recognition (ASR) systems consistently underperform for speakers of African American Vernacular English (AAVE) \cite{koenecke2020racial}. 
While model-level analyses can effectively surface such disparities, they often focus on immediate, quantifiable harms. The evaluations provide only a preliminary look at biases, and harms may manifest differently or propagate more broadly in real-world contexts. 
Additionally, such analyses typically assume clearly defined use cases. This assumption is increasingly untenable in the context of GPAI, where models are deployed across multiple, unforeseen applications.

\paragraph{System Level.} System-level fairness focuses on how AI outputs interact with human decision-makers and organizational processes, particularly where individual-level decisions are made within structured organizations. Unlike model-level analysis, which centers on output disparities, system-level analysis addresses how those outputs shape specific decisions---such as in hiring, lending, or legal judgments---within real-world workflows. For example, ASR models that exhibit linguistic biases may impact decisions that are made based on resulting transcripts, such as when applied to courtroom transcription \cite{prasad2002automatic, martin2023bias}. If transcripts with systematic errors for particular social groups become part of official records, these inaccuracies can compound existing legal disparities, affecting judicial outcomes and access to justice. Similarly, predictive models used for hiring, lending, or law enforcement may inadvertently reproduce or even amplify existing biases within organizational systems~\cite{cohen2025two, langchainKlarnasAssistant}. Identifying system-level harms presents greater methodological complexity than model-level analyses, as these impacts are dependent on both context and human operators of technology. Effective analyses at this level require information-gathering that captures interactions between model outputs and institutional practices, highlighting the need for ongoing, context-sensitive evaluations.

\paragraph{Society Level.}
Society-level fairness examines how the cumulative effects of decisions made with AI in the system reshape broader patterns of inequality across populations. Rather than focusing on individual harms, this level particularly assesses how AI integration influences the long-term distribution of resources, opportunities, and risks. For example, in healthcare, models that underdiagnose disease by applying historical proxies like medical cost data and race-adjusted metrics \cite{seyyed2021underdiagnosis, obermeyer2019dissecting, eneanya2019reconsidering} can compound to lower the quality of care received by marginalized populations overall. In education, AI-based learning tools may be applied to cover gaps in underfunded school systems \cite{AI_underfunded, AI_achievement_gap} despite questionable impact on learning rates \cite{Bastani_Bastani_Sungu_Ge_Kabakcı_Mariman_2024}, resulting in inconsistent quality of education. In labor markets, AI-driven automation and decision-making systems can displace workers or shift them into more precarious positions, intensifying wage suppression and economic insecurity \cite{capraro2024impactgenerativeartificialintelligence}. While all of these issues can individually occur at the system level, society-level fairness asks whether the existence of AI in a particular ecosystem reifies disparities in health, wealth, and power. This broader framing requires an analytical approach that considers how AI systems degrade or enhance resources, who retains access to higher-quality services, and how these dynamics shape broader patterns of inequality over time.
This can involve a variety of work, including economic analyses like the Anthropic Economic Index \cite{anthropic2025economic} that tracks the distribution of benefits and harms as AI becomes integrated into critical sectors, or efforts to evaluate impacts of AI and fairness interventions broader societal wellbeing over time \cite[e.g.,][]{liu2018delayed}. 
Beyond sector-related trends, society-level fairness also entails examining how the energy consumption and carbon emissions of AI systems are distributed \cite{luccioni2024light}. This may disproportionately affect communities with fewer resources and less resilience to climate impacts.
Such assessments are only possible if providers and deployers are transparent about use cases, deployment contexts, and system integration. Accordingly, we emphasize the need for actors to contribute to information-gathering across model, system, and society levels to comprehensively assess the distributional consequences of GPAI systems.

\subsection{Actors and Priorities in AI Regulation}
We delineate the distinct responsibilities of system providers and system deployers. This categorization of actors is particularly relevant in light of regulatory frameworks like the EU Artificial Intelligence Act, which categorizes providers as entities who develop or place AI systems on the market and deployers as those utilizing AI systems under their authority beyond personal, non-professional use \cite{eu2024aiact}. While the AI lifecycle envisions the roles at different stages, entities may assume dual responsibilities depending on their level of control and intervention in the model's development and deployment. 
Importantly, the boundary between these roles can blur when a deployer undertakes significant modifications to a GPAI. Such modifications may assign an entity both deployer and provider status under the AI Act, expanding their obligations. By adopting the same approach as the EU, we underscore the necessity of holding such entities accountable for fairness-related risks, regardless of their initial role in the AI supply chain. This approach ensures a more comprehensive accountability framework, aligning responsibilities with the actual level of influence an entity exerts over a model's outputs and deployment.

Emerging AI regulatory frameworks have lacked specificity on fairness concerns compared to safety, security, and catastrophic risks. Most approaches apply a similar two-actor framework highlighting multiple intervention points, yet existing policies tend to specifically target acute risks visible at the model or system level rather than pervasive society-level harms \cite{bernardi2025societaladaptationadvancedai}. For instance, while the EU AI Act does include non-discrimination provisions, its criteria for high-risk systems emphasize health and safety concerns rather than directly addressing biases and inequities \cite{eu2024aiact}. Similarly, the General-Purpose AI Code of Practice prioritizes mitigating ``serious incidents and malfunctions'' rather than broader, aggregate impacts of biased AI outputs \cite{eu2025gpaicode}. In the U.S., regulatory efforts such as President Trump's Executive Order 13859 emphasized national security, economic competitiveness, and human flourishing without addressing discrimination or bias explicitly \cite{eo14179}, while other orders like Executive Order 13960 \cite{eo13960} and President Biden's Executive Order 14110 \cite[now revoked,][]{eo14110} offered only broad, aspirational references to fairness and civil rights, lacking specific guidance or requirements.

This regulatory gap is further compounded by fairness frameworks that remain largely abstract and decontextualized, emphasizing model-level metrics while overlooking the complex socio-technical dynamics that shape how AI systems interact with institutional processes and user populations \cite{selbst2019sociotechnical, green2018myth}. Given their proximity to deployment settings, deployers are uniquely positioned to surface these emergent risks through localized evaluations and ongoing monitoring. However, a general assessment of social correlations, which developers are suited to provide, remain a helpful heuristic for downstream users and can serve as an incentive for developers to prioritize bias as a relevant model characteristic. Since existing regulatory frameworks provide little guidance or accountability for such interventions, we argue that a multifaceted approach focused on information-gathering by both providers and deployers is essential for detecting and mitigating context-dependent harms.

\subsection{Information-Gathering as a Fairness Intervention}
While information-gathering is often framed as a passive activity, it can also serve as a direct intervention in mitigating fairness harms by surfacing and deterring problematic practices in real time. When system providers and deployers are required to disclose demographic breakdowns of datasets, algorithmic design decisions, or model recipients, they are compelled to confront potential biases and inequities that might otherwise remain obscured. In a handful of examples, companies have altered practices to avoid reputational damage, regulatory scrutiny, or potential litigation. For instance, algorithmic audits that uncover racial disparities in predictive policing systems have led to public outcry, resulting in the suspension of those systems or substantial modifications to their deployment \cite{raji2020closing}. Similarly, well-publicized audits have occasionally led to changes in deployed models~\cite{raji2019actionable}. By making fairness risks publicly visible, systematic data collection can function as a proactive corrective measure, not just a retrospective assessment.

Information-gathering can prompt some action that promotes fairness, but it is no guarantee of accountability; without sufficient context and mechanisms for redress, transparency may not prevent harmful practices and can in fact cause harm if it reveals sensitive information or requires excessive effort to realize benefits \cite{ananny2016seeing}. Thus, we advocate for a measured approach that distributes burdens equitably and responsibly selects which information is relevant. To prescribe actions that mitigate fairness, we must first understand how the different components of GPAI ecosystems interact, such as how pre-training data can propagate into a downstream fine-tuned model. Until such technical understanding is achieved, efforts should center on developing contextual and holistic understandings of fairness-related factors.

\section{Provider Obligations}
When assigning responsibility for fairness-related concerns in AI systems, it is natural to focus on system providers as key stakeholders. The European Union's AI Act, particularly Article 53, outlines general obligations for providers \cite{eu2024aiact}. In this section, we consider a narrower question: whether providers who are \textit{not} also deployers bear affirmative obligations related specifically to fairness. Our position is grounded in the broader recognition that fairness, like safety, is not an intrinsic property of a model itself but emerges through its deployment and use \cite{narayanan2024aisafety}. We argue that, given current understandings of the bias transfer hypothesis (i.e., that bias in pretrained models will propagate into downstream ones) and dual-use cases, providers bear more indirect than direct responsibilities.

One prior regulatory approach has been to eliminate bias from models and data. Early drafts of the EU AI Act proposed that data used in high-risk systems be ``sufficiently relevant, representative and free of errors and complete in view of the intended purpose'' \cite{eu2024aiact}. The final version softens this requirement to mandate that datasets be ``relevant, sufficiently representative, and to the best extent possible, free of errors and complete,'' though even determining what constitutes sufficient representativeness is arguably context-dependent \cite{eu2024aiact}. This shift reflects a broader recognition that social biases are relative to the given context and cannot be eliminated when training general-purpose models. For similar reasons, ``debiasing'' a model poses a challenge. For example, the dual-use nature of AI systems complicates fairness interventions. A model that has been ``de-biased'' to avoid producing racially discriminatory language may no longer be usable for socially valuable tasks, such as identifying racially restrictive covenants in property deeds, legal clauses that historically excluded people of certain races from home ownership or occupancy \cite{surani2024racialcovenants}. However, reporting general correlations between model behaviors and group attributes still helps direct attention and resources toward areas where fairness concerns are most likely to emerge, even when contextual impact remains uncertain.

While system providers should address clear and egregious fairness issues---such as datasets where all depictions of a group are inappropriate or offensive---many harms are more subtle. Tools like model cards and dataset documentation are now common mechanisms for provider transparency \cite{liang2024whatsdocumentedaisystematic}. However, we still lack a robust understanding of how training data and modeling decisions shape fairness outcomes in downstream applications. This gap hinders deployers trying to assess risks and undermines the development of enforceable standards for fairness.

We support prior calls for transparency regarding various social considerations in GPAI \cite{luccioni2024light, ilo_ai_labour_disclosure, bommasani2025fmtindex}, including clarifying environmental harms of model training and deployment and their effect on local communities, ensuring fair labor practices for data workers, and respecting intellectual property rights in the construction of training datasets. While many of these issues are not fairness-specific and relate to more general harms, marginalized communities are often disproportionately affected by these system and society-level harms \cite{gyevnar2025aisafetyeveryone}. This also means that practices that reduce environmental harm, labor exploitation, and power concentration can also mitigate disparities in who is most affected by AI development \cite{hoes2025existential}. In particular, we emphasize that translating these transparency principles into practice does not require exposing proprietary details; instead, reporting relevant, systematic, and structured information can provide an actionable foundation for fairness interventions. For example, reporting regional energy consumption for inference workloads can help identify disproportionate environmental burdens in low-income or climate-vulnerable communities and prompt targeted investment in renewable infrastructure or demand-shifting policies. Over time, as the field matures, deeper insight into the interaction between provider-side decisions and downstream adaptations such as fine-tuning may support more effective governance. We also highlight two additional areas where systematic information-gathering would be particularly valuable.

\paragraph{Evaluating the Impact of Development Decisions.}
System providers, who control key parts of the development pipeline and generally possess substantial resources, must systematically evaluate and invest in research that clarifies how their decisions at different stages affect future model outputs. This is essential for developing actionable fairness interventions and building guardrails that remain effective even as models evolve. In particular, providers should prioritize understanding which biases persist through pretraining and fine-tuning pipelines and how they manifest in model outputs through correlations between demographic attributes and model performance \cite[e.g.][]{kumarDetecting}. They should also examine how fine-tuning can be used to mitigate harmful biases, considering both the quantity and type of tuning to concretely improve models \cite[e.g.,][]{qi2025tokensdeep}. Evaluation should likewise assess and improve the robustness of model-level fairness guardrails \cite[e.g.,][]{qi2024finetuningcompromise, wang2023overwriting}, including through adversarial attempts to produce extremely problematic behavior \cite[e.g.,][]{wallace2025estimatingworstcasefrontierrisks}. A systematic evaluation agenda of this kind not only enables more targeted and effective fairness interventions, but also contributes to the long-term stability of model behavior, fosters broader user adoption, and offers providers a competitive edge. In the event that a provider lacks the resources to make substantive investments in these areas of AI research, they should expand their disclosures accordingly to enable external evaluators to conduct this research.

\paragraph{Disclosing Supply Chain Relationships.}
We argue that an important obligation for providers of GPAI is the disclosure of supply chain relationships. Without visibility into which deployers are using a model and in what domains, it becomes challenging to trace where and how fairness harms emerge and could be addressed. When downstream systems exhibit discriminatory behavior, it is often unclear whether the cause lies in upstream model characteristics, deployment-specific modifications, or contextual misuse. Supply chain disclosures provide the necessary information infrastructure to attribute responsibility appropriately, enabling researchers, regulators, and impacted communities to identify the relevant actors, investigate causal pathways, and design targeted interventions supported by research. In its absence, providers and deployers may deflect accountability onto one another, stalling harm mitigation. Transparency also supports the conditions under which third-party auditors, civil society organizations, and academic researchers can conduct meaningful evaluations, particularly in high-impact domains where fairness harms may otherwise remain obscured.

Beyond attribution and accountability, supply chain disclosures also play a critical role in enabling fairness at scale. When the same base model is reused across a wide array of applications, its embedded biases can replicate and compound across sectors, resulting in algorithmic homogenization~\cite{creel2022leviathan, bommasani2022homogenization}. Individuals may face repeated disadvantages across employment, education, and housing decisions if the same flawed inference patterns follow them from system to system. Disclosing deployment relationships allows auditors to monitor the cumulative effects of model reuse and identifying systemic risks that might be invisible in isolated evaluations. Moreover, such visibility enables anticipatory governance: with sufficient information about where and how models are used, it becomes possible to flag high-risk applications, monitor sensitive domains for overconcentration, and align fairness interventions with actual deployment contexts. While some disclosure obligations may need to be scoped to protect proprietary information, carefully designed transparency regimes—whether through regulators, certification bodies, or consented data sharing—can help balance commercial concerns with the public interest in equitable AI deployment.

\section{Deployer Obligations}

While providers may release context-agnostic metrics displaying social correlations with model performance, the fairness of general-purpose models depends on how they are used in specific contexts. The EU's General-Purpose AI Code of Practice describes the commitments of those involved in the deployment of GPAI systems \cite{eu2025gpaicode}. Specifically, it highlights obligations to formally document and disclose relevant information about models and to assess and mitigate potential harms in AI systems, including issues like illegal discrimination and bias in high-risk application areas.

Because the applications of general-purpose models vary, fairness cannot be guaranteed by focusing on specific social outcomes. A model that appears fair in one use case and level may have harmful effects in another, depending on factors such as the population it affects, the decision-making context, and the way it is integrated into broader systems. As a result, fairness must be embedded into AI deployment through procedures that allow for iterative and context-sensitive measurement and development. System documentation and evaluation thus become critical information-gathering approaches to uphold fairness and engage with the ways a model interacts with particular environments.

\subsection{Disclosing Usage Contexts}
We call for the disclosure of three kinds of information that are critical for ensuring AI fairness. Specifically, information about the social groups that users of AI systems belong to, how the system stores and personalizes responses, and how users are interacting with the system are critical to evaluating and subsequently improving the fairness of AI systems.

\subsubsection{Social Group Labels of Users.} First, we echo calls for more disclosure of the group membership of the users of GPAI systems \cite{bogen2020awareness, ho2020affirmative}. These are needed to determine where a model underperforms so that targeted intervention can help improve overall accuracy. For instance, during machine learning training, it is common practice to search for the ``hardest'' training examples. Collecting group attribute labels can serve as a proxy to help identify subsets of the data that a model is not performing as well on. However, it is not always safe for members of marginalized communities---such as non-citizens and queer communities---to share their identities, and inferring or collecting the data may place the communities at increased risk of surveillance \cite{Tomasev2021Unobserved, ananny2016seeing, wachter2020affinity, Bogen2024NavigatingDemographic}. To safeguard privacy, organizations should aim to follow siloed processes to collect, access, and analyze the data and to minimally determine the attributes of interest \cite{NAIAC2024DataPrivacy, King2023PrivacyBias}.

\subsubsection{Personalization.} Next, we call on deployers to disclose the type of personalization employed by GPAI. It is important to understand how much of a user's interaction history is recorded and used. For example, DeepSeek reportedly collects user interaction data on servers in China, and Snapchat uses user chat history to personalize recommendations \cite{newman2025deepseek, snap2023microsoft}. Released human-LLM datasets have been found to contain personally identifiable information (PII), detailed sexual preferences, and specific drug use habits \cite{mireshghallah2024trustbotdiscoveringpersonal}, which users may not have intended to make public or accessible for personalization purposes. These privacy risks can also include phone numbers and home addresses of individuals. 
Transparency in how personalized advertising is generated with user data has remained limited due to the complexity of cookie-based tracking across multiple online platforms. It is essential to provide clear and accessible information about whether companies retain interaction history, and if so, how it is utilized.

Disclosing how personalization works is critical not only for privacy protection, but also for distinguishing personalization and stereotyping. For instance, there are documented concerns about stereotyping based on characteristics such as a person's name \cite{wilson2024bias}. The release of system prompts in particular can provide insight into how chatbots are instructed to use different types of information and avoid problematic content. For instance, DeepSeek has released their system prompt saying ``Your answers should not include any harmful, unethical, racist, sexist, toxic, dangerous, or illegal content. Please ensure that your responses are socially unbiased and positive in nature'' \cite{deepseekai2024deepseekllmscalingopensource}. Increasing public knowledge of the internal workings of GPAI arguably reduces security risks \cite{Hall_Mundahl_Park_2025}. Similar to GDPR, transparency around personalization should also be dual-layered, containing the type of personalization made public and the specifics accessible to each user.

\subsubsection{Use Cases of AI Systems.} Finally, we echo previous calls for the disclosure of the use cases and supply chain of AI models \cite[e.g.,][]{zhao2024wildchat}. Without information about how AI systems are used in practice, it is hard to understand how to evaluate them and which actors are best positioned to do so. Contexts like summarizing patient records or answering students' questions about academic material \cite[e.g.,][]{handa2025education} can inform the specific tests used to evaluate the model as well as the tolerance level for failures. However, usage-based evaluations are only possible with transparency of use cases. This includes not only knowing how a model is used but also whether it has been modified from its original form, as such changes may significantly affect fairness outcomes in deployment. In 2020, U.S. Executive Order 13960 established the Agency Inventory of AI Use Cases \cite{eo13960}. While this policy underscores the value of cataloging AI use, expanding such disclosure to private companies poses practical and legal challenges, particularly given the difficulties already encountered by federal agencies. Still, targeted disclosure of both use cases and model modifications---such as in high-risk domains or through voluntary standards \cite[e.g.,][]{harvard2023voluntary}---could help support downstream fairness evaluations without imposing excessive burdens.

Overall, from deployers we advocate for three forms of transparency---group labels, personalization levels, and use cases---that can help users make informed decisions about which models to use based on how they may be treated as a result. Group and use-case information particularly illuminates AI's economic impact across occupations and demographics while creating market incentives for responsible governance. These disclosures need not be fully public: demographic data can be reserved for auditors and agencies, and use-case or supply chain information, while potentially sensitive, carries limited consumer risk and may even benefit firms reputationally \cite{Kraft2021Supply}. Still, market incentives alone are insufficient; regulation is needed to ensure companies disclose socially valuable information to relevant stakeholders. Although some business relationships are already subject to disclosure (e.g., SEC filings), they remain difficult to access. Together, these forms of transparency can improve evaluation by enabling auditors to better approximate real-world deployment conditions.

\subsection{Multi-Level Fairness Evaluation}

With adequate transparency about the usage and development of AI systems, it is possible to conduct robust evaluations. These, too, are an information-gathering mechanism that identifies where and how a model produces distinct output distributions, enabling targeted unfairness mitigation strategies and informing responsible deployment. We identify three important design decisions that should be considered when mandating evaluations or scoping models based on the results of evaluation.
These include whether the evaluation is done: (a) before and after model deployment, (b) data that reflects natural and unnatural (i.e., synthetic) contexts, and (c) from within and outside the developer organization. While both sides of each of these axes are valuable, they are not implemented to the same degree and have different implications for the levels of fairness. Thus, we recommend a balanced approach to evaluate fairness at each level.

\begin{table}
\begin{tabular}{@{}llll@{}}
\toprule
\textbf{Evaluation}                           & \textbf{Stage}          & \textbf{Data}      & \textbf{Actor}    \\ \midrule
Benchmarks                           & \cellcolor{stage1}Pre-deploy  & \cellcolor{data1}Synthetic & \cellcolor{actor1}Internal \\
Bug bounty                           & \cellcolor{stage1}Pre-deploy  & \cellcolor{data1}Synthetic & \cellcolor{actor2}External \\
Historical data                      & \cellcolor{stage1}Pre-deploy  & \cellcolor{data2}Natural   & \cellcolor{actor1}Internal \\
Compliance audit                     & \cellcolor{stage1}Pre-deploy  & \cellcolor{data2}Natural   & \cellcolor{actor2}External \\
Simulations & \cellcolor{stage2}Post-deploy & \cellcolor{data1}Synthetic & \cellcolor{actor1}Internal \\
Public red teaming                   & \cellcolor{stage2}Post-deploy & \cellcolor{data1}Synthetic & \cellcolor{actor2}External \\
A/B testing                     & \cellcolor{stage2}Post-deploy & \cellcolor{data2}Natural   & \cellcolor{actor1}Internal \\
Incident reporting                   & \cellcolor{stage2}Post-deploy & \cellcolor{data2}Natural   & \cellcolor{actor2}External \\ \bottomrule
\end{tabular}
\caption{Evaluation categorization by deployment stage, data type, and actor performing the audit. No single evaluation covers all fairness concerns; rather, this typology shows that complementary evaluations are needed to identify and mitigate diverse harms.}
\label{tab:data_categorization}
\end{table}

\subsubsection{Pre-deployment vs. Post-deployment.}
We distinguish between different temporal stages of evaluation, pre-deployment and post-deployment, not as a binary, but as a continuum of iterative oversight. In pre-deployment, an AI system is evaluated before being released to users. This evaluation often prioritizes internal performance standards, typically focusing on technical benchmarks, risk analysis, and compliance with field norms. One common form of pre-deployment evaluation is benchmarking, where suites such as HELM \cite{liang2023holistic} assess models along multiple dimensions including accuracy, robustness, calibration, and fairness. HELM aims to provide a more comprehensive evaluation than traditional leaderboards by including a broad range of scenarios and tasks designed to capture different aspects of model performance, including some fairness-related concerns. Another example is red teaming, in which adversarial inputs are designed to elicit harmful outputs from models \cite{perez-etal-2022-red}. Red teaming efforts often focus on areas like cybersecurity, biosecurity, and content safety, but are generally scoped to harms anticipated by system providers prior to release.

However, pre-deployment evaluations are limited by their reliance on assumptions about user behavior and model usage. Benchmarks are constrained to predefined tasks and may not capture fairness-related harms that arise in diverse real-world settings. Even Dynabench, which introduces dynamic data collection by allowing models to be tested and improved through adversarial user inputs over time, is still limited because it relies on the specific tasks and populations involved in the benchmark construction \cite{kiela-etal-2021-dynabench}. Projects driven by specialized online communities have historically experienced challenges in recruiting marginalized individuals, e.g. Wikipedia editors \cite{hill2013wikipediagendergap}. As a result, it may fail to identify fairness failures that arise in new contexts, affect groups who were not represented among early users or adversaries, or manifest only after prolonged real-world interaction \cite[e.g.,][]{apnews2025chatbotlawsuit}\footnote{Content warning: Discussion of suicide.}. Pre-deployment efforts provide important baselines but are insufficient to ensure ongoing fairness once GPAI is widely used.
Post-deployment evaluation, on the other hand, assesses GPAI based on how users actually interact with it in the real world. This allows evaluation to address discrepancies between intended and actual use, such as differences in the types of questions asked of general-purpose models and reasoning models. Post-deployment evaluation systems can capture emergent user and model behaviors, novel use cases, and biases that would not have been surfaced during pre-deployment testing. 

Different approaches to post-deployment evaluation exist. One is adverse event reporting (AER) systems, which allow users to flag harmful or problematic outputs encountered in practice \cite{naiac2023recommendation}. While AER is crucial for identifying significant incidents, it is both reactive and inherently limited by users' ability to recognize and report harm in the moment. Individuals impacted by pervasive harms may not immediately realize that they have been affected. Incident aggregation efforts such as the \citet{aidb}, which collects and catalogs instances of AI failure, can identify a subset of broader patterns over time, even as they are constrained in scope by the nature of problems that get reported. Such databases offer another layer of analysis by systematically reviewing the totality of user interactions to detect fairness failures at scale \cite{dai2025individualexperiencecollectiveevidence}.

The distinction between pre-deployment and post-deployment evaluation is critical from a fairness perspective because each stage surfaces different types of harm across the model, system, and society levels. Pre-deployment evaluations like benchmarking primarily target model-level disparities, identifying performance gaps in controlled conditions. However, post-deployment evaluations like incident reporting are essential for detecting system-level failures, such as misalignments between outputs and institutional workflows, and society-level harms that emerge over time, like unequal access to services or labor displacement. These broader impacts often affect marginalized groups in ways that cannot be fully anticipated during development. A fairness-centered evaluation pipeline must therefore include both rigorous pre-deployment testing and sustained post-deployment monitoring to capture evolving, context-dependent harms.

\subsubsection{Naturalistic vs. Non-naturalistic Assessment.} 
We define naturalistic assessments as benchmarks that draw from real-world data sources and reflect authentic, ecologically valid scenarios~\cite{de2020towards}. Such assessments are especially critical for system and society-level fairness harm detection because they directly measure the risks, harms, and disparities that models may propagate in practice~\citep{shen2025societal}. In contrast, non-naturalistic assessments that rely on synthetic data generation or controlled perturbations are generally limited to detecting model-level harms, such as output disparities across demographic attributes. While these evaluations provide important insights into model behavior under controlled manipulations, they often lack the contextual richness necessary to capture systematic and structural biases. Prior work has shown that simple perturbations, such as switching demographic attributes in text, can produce absurd or misleading results if real-world social dynamics are not considered. \citet{blodgett-etal-2021-stereotyping} famously illustrate this with the ``Norwegian salmon" example, where perturbation-based measures of stereotyping confound nationality and race, leading to spurious conclusions. Naturalistic evaluation guards against such failures by grounding fairness measures in the actual ways that identity, language, and power interact in real environments.

Beyond sociolinguistic tasks, the importance of ecological validity has been highlighted in high-stakes domains like cybersecurity. For instance, the BountyBench framework evaluates language models on three common security tasks: detecting, exploiting, and patching security vulnerabilities. It uses bug bounties rather than synthetic tasks constructed by researchers \cite{zhang2025bountybenchdollarimpactai}. The findings demonstrate that realistic, complex tasks expose significant model limitations and vulnerabilities, which synthetic researcher-constructed datasets might not. The researchers also attach monetary values to tasks to approximate economic impacts. Thus, authentic benchmarks better reflect the multifaceted, emergent risks and capabilities of AI systems, and metrics should be grounded not just on model outputs but also in approximations of impact and human-level baselines, using historical human data or controlled trials to contextualize what fair and reliable performance should look like.

That said, non-naturalistic data remains essential for evaluating fairness failures at the model level. Controlled perturbations and synthetic counterfactuals allow for fine-grained analysis of specific model behaviors that might be difficult to isolate in naturally occurring data. For instance, counterfactually augmented datasets help models learn to distinguish spurious correlations from causal signals \cite{kaushik2020learningdifferencemakesdifference}. Although such synthetic interventions cannot substitute for real-world grounding, they provide critical tools for stress-testing models and diagnosing fairness failures at a granular level. In a comprehensive fairness evaluation framework, naturalistic assessments should anchor primary evaluations of harm and risk, while non-naturalistic methods should serve as supplementary diagnostics for bias and robustness.

\subsubsection{Internal vs. External Evaluation.} It is critical to consider the position and incentives of those conducting evaluations when assessing fairness in AI systems. Internal evaluation plays an important role, particularly due to the greater access internal auditors have to models, data, and development teams. Internal teams can conduct more detailed audits, identify model-level fairness failures and system-level repercussions early, and work closely with providers to prioritize fixes \cite{raji2020closing}. For example, OpenAI conducted an internal audit examining biases in chatbot interactions by analyzing responses to users of different backgrounds. This study would only be possible through access to sensitive internal data \cite{eloundou2024first}. Anthropic has also analyzed economic implications and values \cite{anthropic2025economic, huang2025values}, which provided insight into fairness issues at the interaction level and informed mitigation strategies. However, internal evaluations remain constrained by organizational priorities, incentives, and perspectives, and thus cannot substitute for independent scrutiny.

External evaluators are essential for uncovering issues that might otherwise be deprioritized or overlooked by model development organizations. These differences can occur because external evaluators may lack company-internal assumptions about the users of the AI and what they will do with it. In particular, external evaluation helps surface fairness concerns that may conflict with profit incentives or internal narratives \cite{longpre2025inhouseevaluationenoughrobust}. Third-party auditing, where independent actors assess models for biases and misalignments with societal values, serves as a key mechanism for achieving public accountability. However, external evaluators often face limited access to proprietary data and model internals, making it difficult to conduct comprehensive assessments. We echo calls to support external evaluation ecosystems, including providing broader access and legal protections to auditors, to enable evaluations that capture the levels of fairness more expansively \cite{raji2022outsider}. Such support is necessary to ensure that diverse stakeholders, particularly those most impacted, have channels for oversight.

Building a robust third-party audit ecosystem is central to enabling effective external evaluations. \citet{raji2022outsider} highlight several critical elements: providing auditors with adequate access to system artifacts, establishing standardized auditing frameworks, and creating safe harbor protections for auditors to mitigate legal risks. \citet{costanzachock2022who} similarly recommend resourcing external auditors and mandating public disclosures of audit results, arguing that without independent scrutiny, algorithmic harms—particularly system and societal impacts affecting marginalized communities—are less likely to be surfaced. External evaluation, when properly supported, ensures that fairness concerns at the system and society levels are evaluated from outside the narrow lens of the organizations deploying these systems.

\subsubsection{From Evaluation to Impact}
We raise these three dimensions of evaluation because they highlight important differences between what information is produced and what conclusions can be drawn about fairness at different levels. For instance, while post-deployment evaluation effectively captures risks that arise in practice, pre-deployment evaluation is critical to ensure that any deployed application has been tested before it is used on real people. At the same time, while naturalistic data is an ideal, using it comes with privacy concerns \cite{mireshghallah2024trustbotdiscoveringpersonal} and issues of scale that synthetic data can help to alleviate. Finally, without the right incentive structures for external evaluators, internal evaluators may currently have the most motivation and access to properly evaluate their models. Regulations for GPAI fairness should therefore encourage complementary forms of evaluation that collectively span the range of possible risks and contexts.

At the same time, even evaluations that look the same according to our distinctions can differ in critical ways. One evaluation method that is post-deployment, naturalistic, and external is adverse event reporting (AER), in which users can report incidents of harmful behavior. However, AER fails to capture more subtle, pervasive fairness issues like erasure~\cite{Katzman_Wang_Scheuerman_Blodgett_Laird_Wallach_Barocas_2023} that cause harm systematically. However, audits based on open-source historical data are also post-deployment, naturalistic, and external but can identify instances of erasure because the data is not pre-filtered for obvious harm.

Ultimately, broader evaluations that span multiple methods and dimensions are not just more informative: they can also drive accountability and change. For instance, when incidents reported through AER require the deployer to implement fixes, this creates a feedback loop that can help deployers and providers to prevent future harm. But more systematic issues in how individuals from different groups are treated demand more proactive forms of oversight. These types of harms often go unnoticed unless surfaced through targeted audits or retrospective analyses, so AER can benefit from being paired with other evaluations. Public-facing audits that produce fairness or safety scores could further shift incentives by enabling users to choose services aligned with their values, generating competitive pressure on companies to prioritize equitable behavior. In this way, a more expansive and layered evaluation ecosystem not only reveals where systems fall short but also creates structural levers for improving fairness in GPAI over time.

\section{Operationalizing Fairness for Regulatory Contexts}
Effective regulation must be scoped to capture the systems most relevant to its goals while avoiding burdensome overreach on smaller-scale deployers \cite{laufer2025backfiringeffectweakai}. Poorly scoped fairness regulation risks reinforcing incumbency by making compliance disproportionately easier for large providers. To avoid this, regulators must attend not only to what is regulated, but how scope is defined. We offer two core recommendations for defining what counts as ``in scope'' for regulation: scope should be determined along multiple dimensions rather than a single threshold, and regulation should apply to systems, not just models. Scoping will necessarily vary depending on whether the regulation targets model developers (where compute might matter more) or deployers (where user scale may be more salient).

Single-threshold rules are insufficient for regulating fairness in GPAI. While compute-based thresholds may serve as rough proxies for certain catastrophic or security-related risks, as applied in President Biden's Executive Order 14110 \cite{eo14110}, they fail to capture the socio-technical mechanisms through which fairness harms arise. These harms occur across the full range of model sizes and capabilities, depending not on compute but on context; even simple models, such as logistic regressions, can produce severe discriminatory outcomes when used in high-stakes areas like credit or hiring \cite{hooker2024limitationscomputethresholdsgovernance}. Similarly, low-compute systems deployed in high-stakes settings, like organ transplant allocation, may still warrant intense scrutiny because of the potential severity of their outcomes, lack of voluntariness of exposure, and disproportionate impacts on vulnerable groups \cite{Hasjim2024Agent}. Assessing potential fairness risks thus requires a contextual, multi-dimensional approach that implies a system-level focus rather than a purely model-centric view.

A system-level lens is necessary because fairness harms emerge from interactions among models, data pipelines, institutional processes, and human oversight. Regulatory experience supports this view: New York City's Local Law 144, for instance, exempted tools with human oversight, yet empirical studies show that such oversight often fails to prevent bias \cite{groves2024auditing, green2022flaws}. Evaluating only the technical artifact ignores how systems actually shape outcomes in practice. We identify dimensions that are most relevant to fairness, which are each dependent on the system as a whole: severity, voluntariness, scale, and distribution of harm. Severity depends on the decisions the system influences; voluntariness on whether individuals can avoid or contest its use; scale on how widely the system is deployed; and distribution on which groups bear the resulting burdens. Once risk is defined along these context-sensitive dimensions, the system naturally becomes the appropriate unit of regulation.

A multi-dimensional, system-based framework also enables proportionate oversight. When risk is assessed in context, regulatory expectations can scale with both the magnitude of potential harm and the resources of the responsible entity. Large providers that command extensive compute and labor capacity can be expected to conduct more comprehensive evaluations, while smaller deployers should face requirements proportionate to their operational scope. Such calibration is necessary to maintain regulatory feasibility. Dynamic thresholds that evolve with empirical evidence can further ensure flexibility, avoiding the procedural rigidity that has hindered other governance regimes \cite{epstein2018many}. In this sense, assessing risk along multiple, context-sensitive dimensions requires a system-level view.

\section{Conclusion}
In this work, we examine what fairness requires in the context of general-purpose AI (GPAI), where the use cases are often undefined at deployment time. While fairness has long been recognized as context-dependent, this raises the question: what obligations are appropriate when the context is unknown? We propose a set of provider and deployer responsibilities grounded in the current technical capabilities and limitations of GPAI. As such, we do not currently call for disclosures like pretraining data or fine-tuning strategies, both because these are often considered proprietary and because, at present, there is limited methodological clarity on how to meaningfully interpret such information. Future research may establish these disclosures as necessary, but that threshold has not yet been met. Instead, we focus on actionable areas where meaningful oversight is currently possible, such as use case categories, worker conditions, and personalization mechanisms. While maximal disclosure may seem appealing, imposing broad requirements risks burdening smaller actors and further concentrating regulatory power in the hands of large providers and deployers.

Because GPAI lacks a predefined context, we cannot set outcome-based prescriptions that we might in domain-specific legislation, such as specific thresholds on predictive performance disparities \cite{watkins2024fourfifths}. Instead, regulation must focus on process: strategically gathering information and prioritizing current harms to build institutional resilience \cite{narayanan2023normal}. We describe the kinds of disclosure and evaluations that are most important, rather than establishing pre-deployment evaluation criteria or benchmarks. Ultimately, we emphasize information gathering to work toward contextual fairness in ostensibly acontextual GPAI systems, arguing that regulation, research, and resources should center on mitigating fairness-related harms so that it becomes possible to leverage AI for the benefit of everyone.

\section{Adverse Impact}

This work offers a focused intervention into fairness-related harms in general-purpose AI, but several limitations and boundary choices are worth acknowledging. First, our regulatory analysis primarily engages with U.S. and European frameworks, such as the EU AI Act and various American executive orders. While these jurisdictions currently shape much of the global conversation around AI governance, our framework does not fully account for legal, cultural, or infrastructural differences in other regions, including Global Majority countries. Future work should examine how fairness responsibilities and data governance norms operate under alternate political conditions and institutional capacities. Second, while we center fairness—particularly in its representational, allocative, and systematic forms—we do not imply that harms like existential risk, misinformation, and ecological sustainability are less important. Our choice reflects a commitment to depth over breadth, with the hope that complementary work will address these parallel challenges.

We also want to emphasize that our call to broaden responsibility beyond system providers is not an attempt to absolve them of accountability for harms stemming from the models they build and distribute. Providers often retain significant control and resources and should be held to strong standards for transparency and impact mitigation. Our argument is that fairness harms frequently emerge from interactions between models and deployment contexts, and that meaningful redress requires shared responsibility. At the same time, we recognize that our proposals—particularly around information-gathering—carry ethical risks of their own. In contexts of weak oversight or coercive governance, increased demands for transparency could be weaponized to deepen surveillance or chill user expression. Similarly, overly burdensome compliance regimes may entrench the power of large incumbents and stifle innovation from smaller actors. Accordingly, any fairness regulation must be paired with robust privacy safeguards, rights-based governance, and attention to how regulatory burdens are distributed. These concerns underscore the need for pluralistic, power-aware implementation and continued reflexivity in how fairness frameworks are operationalized.

\section{Positionality Statement}

Our team draws primarily from backgrounds in computer science and law and well-resourced institutions in the United States. As a result, our analysis may prioritize technically feasible interventions that are actionable within current regulatory and development pipelines. However, we recognize that our perspectives are shaped by our positions within the Global North. As such, we may underemphasize lived experience, localized concerns, or political economies of harm outside U.S. and European regulatory contexts.

\section{Acknowledgements}

JHS is supported by the Simons Foundation Collaboration on the Theory of Algorithmic Fairness and the Simons Foundation investigators award 689988. 






\bibliography{references}

\end{document}